\documentclass[aps,prc,showpacs, showkeys,floatfix,footinbib,twocolumn,superscriptaddress,10pt]{revtex4-1}

\usepackage{amsmath}    
\usepackage{graphicx}   
\usepackage{verbatim}
\usepackage{color}      
\usepackage{hyperref}   

\usepackage{mathptmx}

\long\def\/*#1*/{}

\begin{document}


\title{Estimation of attractive and repulsive interactions from the fluctuation
  observables at RHIC using van der Waals hadron resonance gas Model}

\author{Subhasis Samanta}
\email{subhasis.samant@gmail.com}
\affiliation{School of Physical Sciences, National Institute of Science 
Education and Research, HBNI, Jatni - 752050, India}

\author{Bedangadas Mohanty}
\email{bedanga@niser.ac.in}
\affiliation{School of Physical Sciences, National Institute of Science 
Education and Research, HBNI, Jatni - 752050, India}
\affiliation{Experimental
Physics Department, CERN, CH-1211, Geneva 23, Switzerland}

\begin{abstract}
Experimental data on the moments of net-proton distribution in central 
Au-Au collisions for various center of mass energies 
 ($\sqrt{\mathrm {s_{NN}}}$) measured by the STAR collaboration 
at the Relativistic Heavy-Ion Collider (RHIC) are compared to the 
corresponding results from a van der Waals type interacting hadron 
resonance gas (VDWHRG) model. The parameters representing the attractive and repulsive interactions 
in the VDWHRG model have been extracted by fitting the $\sigma^{2}$/$M$,
$\it{S}\sigma$ and $\kappa\sigma^{2}$, where $M$ is the mean, $\sigma$
is the standard deviation, $\it{S}$ is the skewness and $\kappa$ the 
kurtosis of the net-proton distribution. Considering all the three
moment products we observe that the strength of the repulsive
interactions increases with decrease in $\sqrt{\mathrm {s_{NN}}}$ = 200 to 19.6
GeV while the strength of the attractive interaction is of the similar magnitude. For $\sqrt{\mathrm {s_{NN}}}$ = 11.5 and 7.7 GeV
there is drop in the strength of both attractive and repulsive
interactions relative to $\sqrt{\mathrm {s_{NN}}}$ = 19.6 GeV. On the
other hand, if we consider only the higher order moment products,
$\it{S}\sigma$ and $\kappa\sigma^{2}$, which are more sensitive to
critical point physics, a segregation with respect to the strength of the attractive parameter is
observed. The data for $\sqrt{\mathrm {s_{NN}}}$ = 19.6 and 27 GeV
supports a larger attractive strength compared to other
energies. For this latter case, the repulsive interaction values are of
similar order for most of the beam energies studied, except for 7.7
GeV where the parameter value is not well constrained due to large
uncertainties. 
\end{abstract}

\maketitle

\section{Introduction}
One of the major goals of high energy heavy-ion collisions
is to explore the phase diagram of strongly interacting
nuclear matter at high temperature and density.
At large temperature $T$
and zero baryon chemical potential $\mu_B$, the lattice quantum chromodynamics (LQCD)
predicts a smooth cross-over transition from hadronic to a quark gluon plasma (QGP) phase~\cite{Aoki:2006we}. 
While at large $\mu_B$ a first order phase transition is 
expected~\cite{Asakawa:1989bq, Alford:1997zt, Ejiri:2008xt, Bowman:2008kc}.
Therefore, there should be a critical point (CP), the end point of the
first order phase boundary towards the crossover 
region~\cite{Fodor:2004nz, Gavai:2004sd, Stephanov:2004wx}.
The beam energy scan (BES) program is currently ongoing at the Relativistic Heavy-Ion Collider
(RHIC) facility at Brookhaven National Laboratory to locate the QCD CP.

Fluctuations of conserved charges in heavy-ion collisions like baryon number, 
electric charge, strangeness quantum number are sensitive observables
for the CP search~\cite{Stephanov:1999zu, Stephanov:2008qz, Stephanov:2011pb, Gupta:2011wh}. 
The non-monotonic variation of these quantities with the colliding 
beam energy is regarded as one of the characteristic signature in
presence of the CP. The STAR collaboration at RHIC has measured the fluctuation 
observables related to the net-proton (a proxy for net-baryon), net-charge and 
net-kaon (a proxy for net-strangeness) distribution~\cite{Adamczyk:2013dal,
  Adamczyk:2014fia, Adamczyk:2017wsl}. 
The product of $\it{S}\sigma$ and $\kappa\sigma^{2}$ for the measured net-proton 
distribution in central Au-Au collisions qualitatively show non-monotonic dependence on the beam 
energy.  Since uncertainties in the measurements at lower 
$\sqrt{\mathrm {s_{NN}}}$ are large, the evidence for the existence of   a CP is not yet conclusive. 
The product of moments of the net-number 
  distributions are also related to the susceptibilities of conserved charges which can be computed in 
LQCD~\cite{Gavai:2008zr,Cheng:2008zh, Gupta:2011wh,Bazavov:2012jq, Bazavov:2012vg, Borsanyi:2013hza} and in the 
hadron resonance gas (HRG) models~\cite{Karsch:2010ck, Garg:2013ata, Fu:2013gga, 
Bhattacharyya:2013oya, Alba:2014eba, Adak:2016jtk, Bluhm:2018aei,Bellwied:2018tkc}.

HRG model has several varieties.
Some of the HRG models consider interaction among the constituent
hadrons and some do not.
Different versions of this model and some of the recent work using 
these models
may be found in Refs. \cite{Hagedorn:1980kb, Rischke:1991ke,Venugopalan:1992hy,
Cleymans:1992jz, BraunMunzinger:1994xr, Cleymans:1996cd, Yen:1997rv, BraunMunzinger:1999qy,
Cleymans:1999st, BraunMunzinger:2001ip, BraunMunzinger:2003zd, Karsch:2003zq, Tawfik:2004sw,
Becattini:2005xt, Andronic:2005yp, Andronic:2008gu,Begun:2012rf, Andronic:2012ut,
Tiwari:2011km,Wiranata:2013oaa, Fu:2013gga, Tawfik:2013eua, Garg:2013ata, Bhattacharyya:2013oya,
Bhattacharyya:2015zka,Chatterjee:2013yga,Chatterjee:2014ysa,Chatterjee:2014lfa,Becattini:2012xb,Bugaev:2013sfa,Petran:2013lja,
Vovchenko:2014pka,Bhattacharyya:2015pra, Kapusta:2016kpq,Begun:2016cva,Adak:2016jtk,
Xu:2016skm,Fu:2016baf,
Vovchenko:2015xja, Vovchenko:2015vxa, Vovchenko:2015pya,Broniowski:2015oha,Vovchenko:2015idt,
Redlich:2016dpb,Vovchenko:2016rkn,Alba:2016fku,
Samanta:2017ohm,Sarkar:2017ijd,Bhattacharyya:2017gwt,Chatterjee:2017yhp,Alba:2016hwx,
Vovchenko:2017zpj, Vovchenko:2017cbu,Dash:2018can, Dash:2018mep}.
The HRG model is successful in describing the zero chemical potential LQCD data on bulk
properties of the QCD matter up to temperatures $T \sim 150$ MeV
~\cite{Borsanyi:2011sw, Bazavov:2012jq, Bazavov:2014pvz,Bellwied:2013cta, Bellwied:2015lba}.
This model is also successful in describing the hadron yields,
created in central heavy ion collisions at different $\sqrt{\mathrm {s_{NN}}}$
~\cite{BraunMunzinger:1994xr, Cleymans:1996cd, Cleymans:1999st, 
Andronic:2005yp,Adamczyk:2017iwn}. Recently, van der Waals (VDW) type interaction with both attractive and repulsive
parts have been introduced in HRG model ~\cite{Vovchenko:2015xja, Vovchenko:2015vxa, Vovchenko:2015pya, Redlich:2016dpb, 
Vovchenko:2016rkn, Samanta:2017yhh}. This model predicts a first order
liquid-gas phase transition along with a CP. Further, this 
model can describe Lattice QCD data of different thermodynamical quantities
satisfactorily. 
For the study of fluctuation of conserved charges,
the HRG model is generally used to obtain a 
baseline for CP search in the experiments. In this regard, the VDWHRG model
may bring in additional dimensions, as it has the capability to
capture both the attractive interactions (important for CP physics)
and the repulsive interactions due to finite size of the hadrons. 

In the present work, we have extracted attractive and repulsive
parameters in the VDWHRG model by fitting the experimental data on
net-proton number fluctuation measured by the STAR collaboration in
BES program at RHIC to the corresponding model results. This then allows for
the first time to get an estimate of the changes in the relative
contributions of the strength of repulsive and attractive interactions
with $\sqrt{\mathrm {s_{NN}}}$ in Au-Au collisions at RHIC.

The paper is organized in the following manner. In the next section we
discuss ideal (non interacting) HRG and the VDWHRG models.
In Sec.~\ref{Sc.Fluctuation}, we discuss the experimental observables
related to the fluctuation of conserved charge. 
Then in the Sec.~\ref{Sc:Extraction} we
discuss the methodology to extract van der Waals parameters of the VDWHRG model 
and the results. Finally in Sec. \ref{Sc:Summary} we summaries our
findings.

\section{Model}
In this section we will briefly discuss the ideal HRG (IHRG) and the VDWHRG models.

\subsection{IHRG model}
In the ideal HRG model, the thermal system consists of non-interacting
points like hadrons and resonances.
The logarithm of the partition function of a hadron resonance gas 
in the grand canonical ensemble can be written as 
\begin {equation}
 \ln Z^{id}=\sum_i \ln Z_i^{id},
\end{equation}
where the sum is over all the hadrons and resonances. 
$id$ refers to ideal {\it i.e.}, non-interacting HRG model.
For particle species $i$,
\begin{equation}\label{Eq:partition_fn}
 \ln Z_i^{id}=\pm \frac{Vg_i}{2\pi^2}\int_0^\infty p^2\,dp \ln[1\pm\exp(-(E_i-\mu_i)/T)],
\end{equation}
where $V$ is the volume of the system,
$g_i$ is the degeneracy, $E_i = \sqrt{p^2 + m_i^2}$
is the single particle energy, $m_i$ is the mass of the particle
and $\mu_i=B_i\mu_B+S_i\mu_S+Q_i\mu_Q$ is the
chemical potential. The $B_i,S_i,Q_i$ are respectively
the baryon number, strangeness and electric charge of the particle,
$\mu^,s$ are the corresponding chemical potentials.
The upper and lower signs of $\pm$ in Eq.~\ref{Eq:partition_fn}
correspond to fermions and bosons, respectively.
We have incorporated all the hadrons and resonances listed in the particle 
data book up to a mass of 3 GeV~\cite{Patrignani:2016xqp}.
Once we know the partition function of the system
we can calculate other thermodynamic quantities. 
The pressure is related to the partition function by the following relation:

\begin{align}\label{eq:p}
  \begin{split}
   p^{id}&=\sum_i \frac{T}{V}\ln Z_i^{id}\\
   &=\sum_i (\pm)\frac{g_iT}{2\pi^2}\int_0^\infty p^2\,dp \ln[1\pm\exp(-(E_i-\mu_i)/T)],
  \end{split}
\end{align}

The $\mathrm {n^{th}}$ order susceptibility can be calculated once we know the pressure
using the formula,
\begin{equation}\label{Eq:chi}
 \chi^n_q = \frac{\partial^n {((p/T^4))}}{\partial {(\frac{\mu_q}{T})}^n}
\end{equation}
where $\mu_q$ is the chemical potential for the charge $q$
which can be any conserved quantity like
B (baryon), S (strangeness), and Q (electric charge) etc.

\subsection{VDWHRG model}
In IHRG model, hadrons and resonances are point-like non-interacting particles.
However, interaction is needed especially at very high temperature or chemical
potential where ideal gas assumption becomes inadequate. Further
hadrons physically have finite sizes. To catch the basic
qualitative features of a strongly interacting system of gas of hadrons, van der Waals type
interaction is incorporated in the VDWHRG model.
The van der Waals equation in the canonical ensemble is given by
\cite{Book_griener}
\begin{equation}\label{eq:vanderWaals}
 \left(p + \left(\frac{N}{V}\right)^2 a \right) (V-Nb) = NT,
\end{equation}
where $p$ is the pressure of the system, $N$ is the number of particles
and $a,b$ (both positive) are the van der Waals parameters. The parameters
$a$ and $b$ describe the attractive and repulsive interaction 
respectively. Higher the value of $a$, the greater the attraction between hadrons
and more probable for a phase transition.
The Eq. \ref{eq:vanderWaals} can be written as

\begin{equation}\label{eq:p}
 p(T,n) = \frac{NT}{V-bN} -a\left(\frac{N}{V}\right)^2 \equiv \frac{nT}{1-bn} - an^2,
\end{equation}
where $n \equiv N/V$ is the number density of particles.
The first term in the right-hand side of the Eq. \ref{eq:p}
corresponds to the excluded volume correction where
the system volume is replaced by the available volume $V_{av} = V -bN$,
where $b = \frac{16}{3} \pi r^3$ is the proper volume of particles with $r$
being corresponding hard sphere radius of the particle.
The second term in Eq. \ref{eq:p} corresponds to the attractive interaction between particles.
The importance of van der Waals equation is that this analytical model
can describe first-order liquid-gas phase transition of a real gas
which ends at the critical point. 
Such a feature is also an expectation for the QCD phase diagram at
larger $\mathrm {\mu_{B}}$ which corresponds to lower  $\sqrt{\mathrm
  {s_{NN}}}$ in the experiments.

The van der Waals equation of state in the Grand canonical ensemble
can be written as \cite{Vovchenko:2015xja, Vovchenko:2015vxa}
\begin{equation}\label{Eq:p_VWD}
 p(T,\mu) = p^{id}(T,\mu^*) - an^2, ~~ \mu^* = \mu - bp(T,\mu) -abn^2 + 2an,
\end{equation}
where $n \equiv n(T,\mu)$ is the particle number density of the van der Waals gas:
\begin{equation}
n \equiv n(T,\mu) \equiv \left(\frac{\partial p}{\partial \mu}\right)_T = \frac{n^{id}(T,\mu^*)}{1 + bn^{id}(T,\mu^*)}.
\end{equation}
Susceptibilities in the VDWHRG model can be calculated by putting pressure 
of Eq.~\ref{Eq:p_VWD} into the Eq.~\ref{Eq:chi}.
In the VDWHRG model, the interactions exist between all pairs of
baryons and all pair of anti-baryons only. The mesons are non-interacting
in this model.

Several methods have been proposed to fix the van der Waals parameters
in VDWHRG model. 
In the Ref. ~\cite{Vovchenko:2015vxa} the VDW parameters $a$ and $b$ (or $r$) have 
been fixed by reproducing the saturation density $n_0 = 0.16$ fm$^{−3}$ and
binding energy $E/N = $ 16 MeV of
the ground state of nuclear matter. The parameters thus obtained in this method
are $a = 329$ MeV fm$^3$ and $r = 0.59$ fm~\cite{Vovchenko:2015vxa}.
The model predicts a first order liquid-gas phase transition which has a CP
at $T = 19.7$ GeV and $\mu_B = 908$ MeV~\cite{Vovchenko:2015vxa}.
While in Ref.~\cite{Samanta:2017yhh}, $a$ and $r$ have been fixed 
by fitting the LQCD data at zero chemical potential. The values of the VDW
parameters in this model are $a = 1250 \pm 150$ MeV 
fm$^3$ and $r = 0.7 \pm 0.05$ fm. A liquid-gas phase transition is observed in this 
model as well with a CP at $T = 62.1$ MeV and $\mu_B = 708$ MeV.

\begin{figure*}
\centering
 \includegraphics[width=0.32\textwidth]{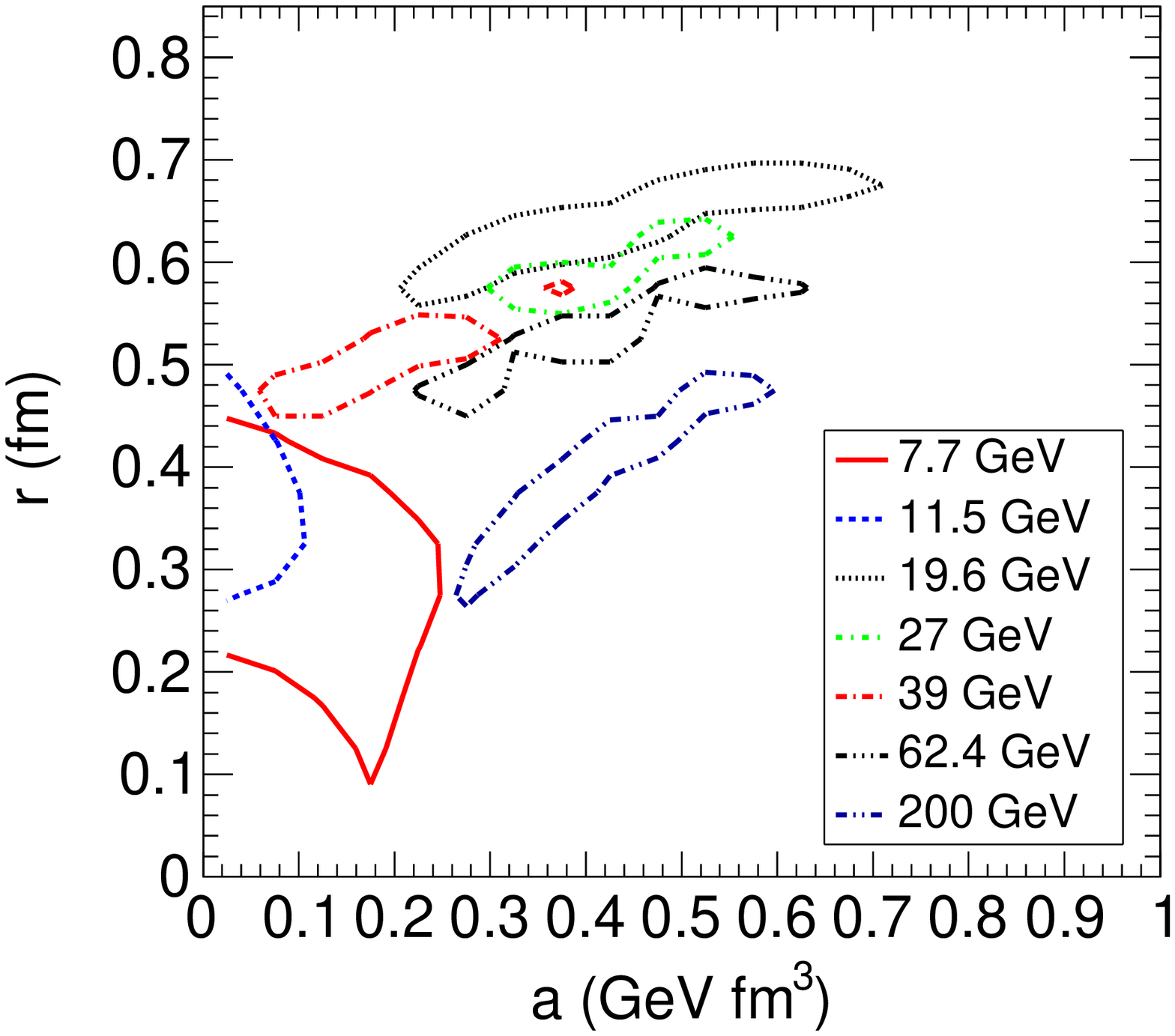}
 \includegraphics[width=0.32\textwidth]{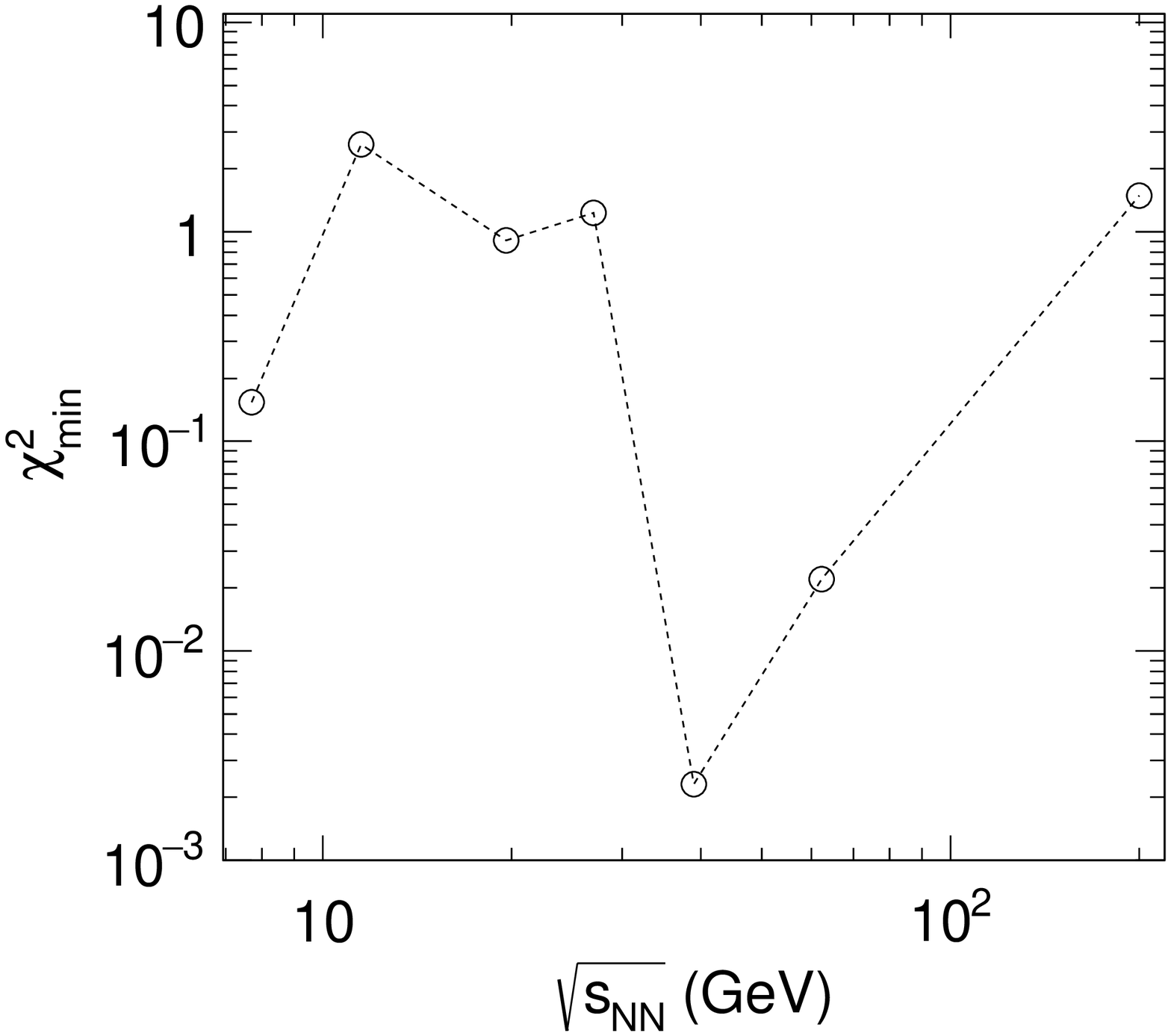}
 \caption{(Left) $1 \sigma$ contours of the 
 van der Waals parameters in $r, a$ plane, at different $\sqrt{\mathrm {s_{NN}}}$ 
extracted from the experimental data of
$\sigma^2/M$, $S\sigma$ and $\kappa\sigma^2$ of net-proton
(VDW parameter Set-1),
(Right) minima of the $\chi^2$ at different $\sqrt{\mathrm {s_{NN}}}$.}
\label{fig:a_r_all}
 \end{figure*}

\begin{table}
\centering
\begin{tabular}{|c|c|}
\hline
VDW parameter &Experimental data used\\
\hline
Set-1 &$\sigma^2/M, S \sigma, \kappa \sigma^2$ \\
Set-2 &$ S \sigma, \kappa \sigma^2$ \\
\hline
\end{tabular}
\caption{Sets of van der Waals parameters and the corresponding experimental
data of net-proton fluctuation used.}
\label{tableLabel1}
\end{table}

\section{Fluctuation observables}\label{Sc.Fluctuation}

\begin{figure*}
\centering
 \includegraphics[width=0.32\textwidth]{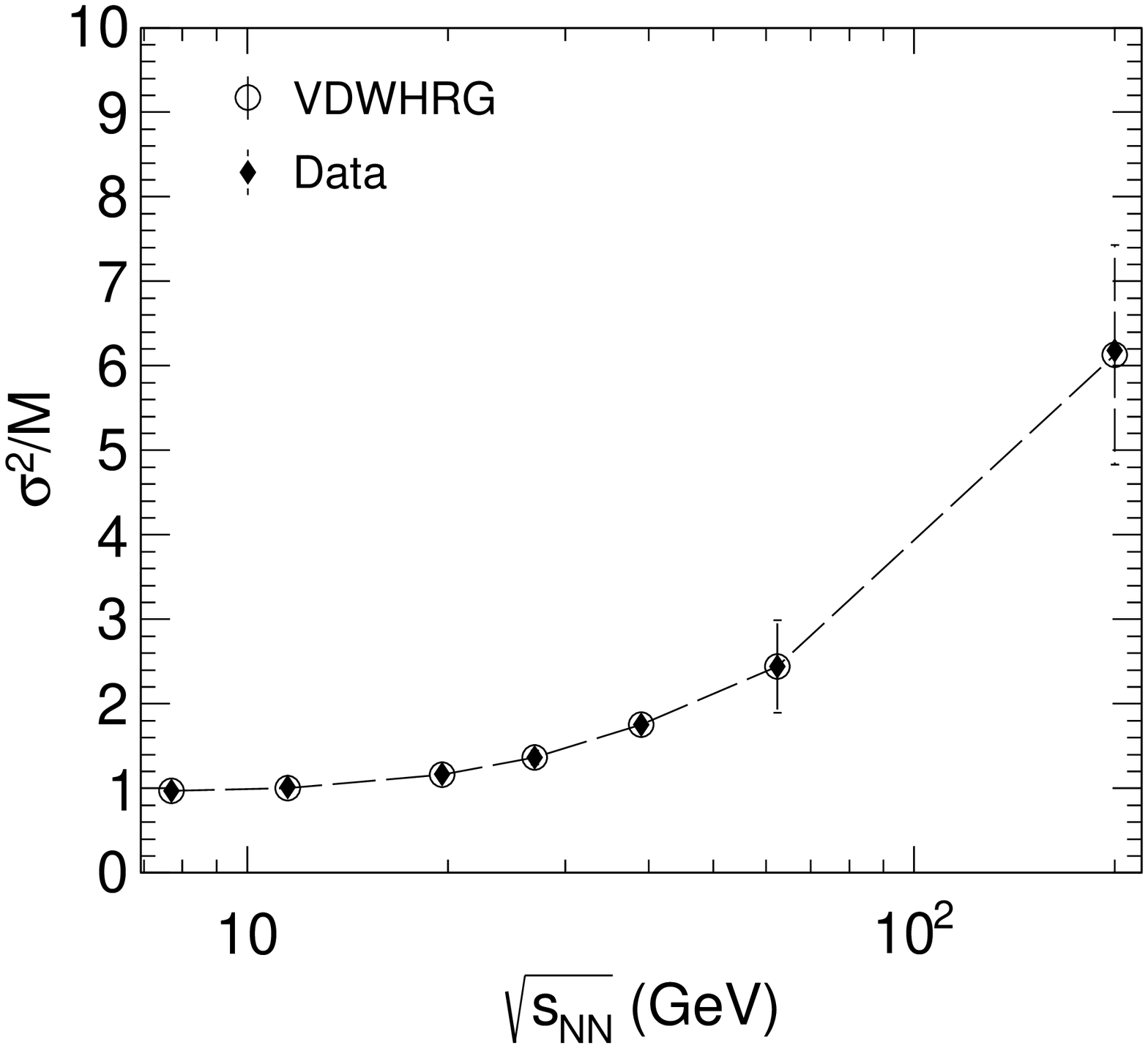}
 \includegraphics[width=0.32\textwidth]{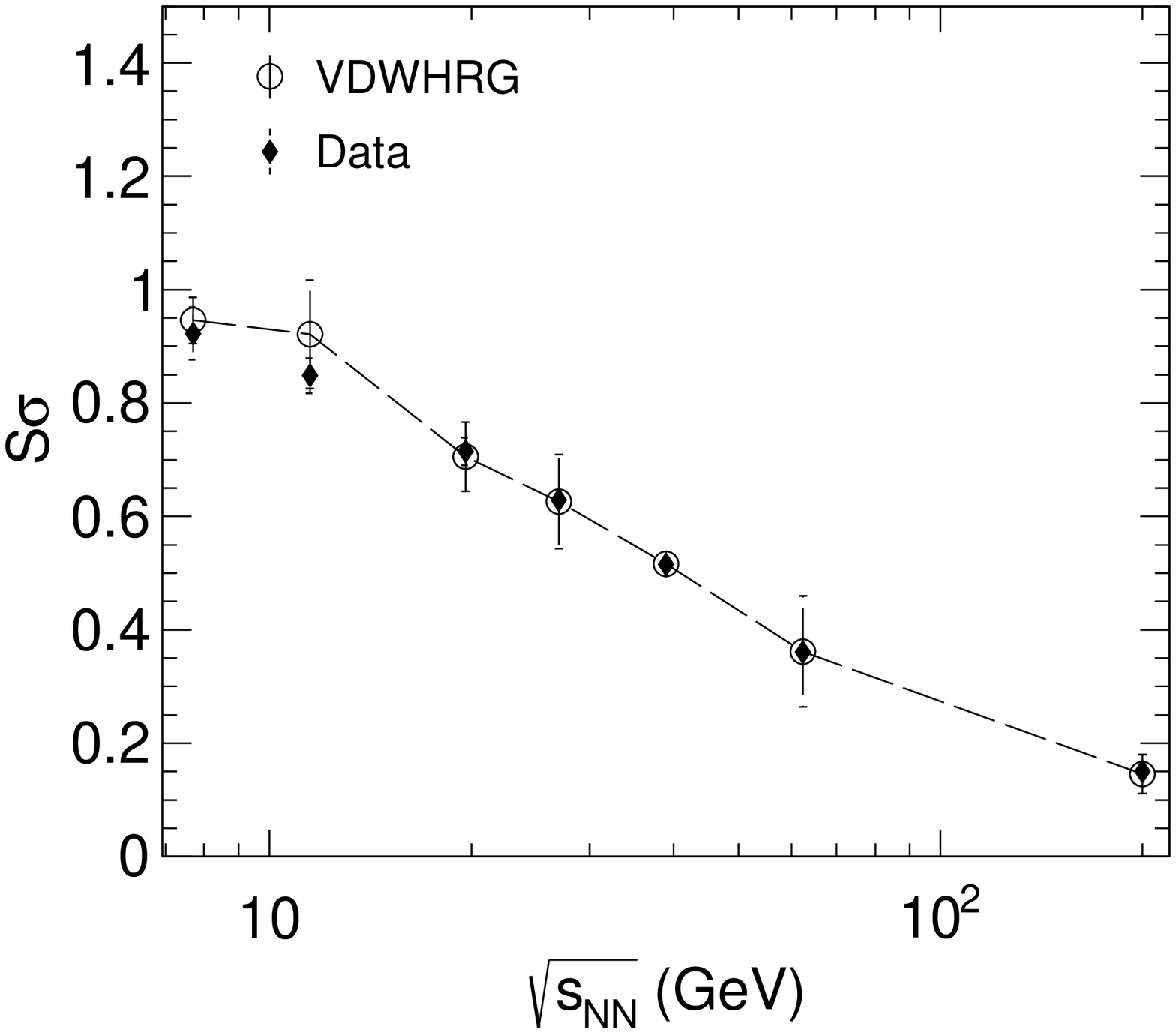}
 \includegraphics[width=0.32\textwidth]{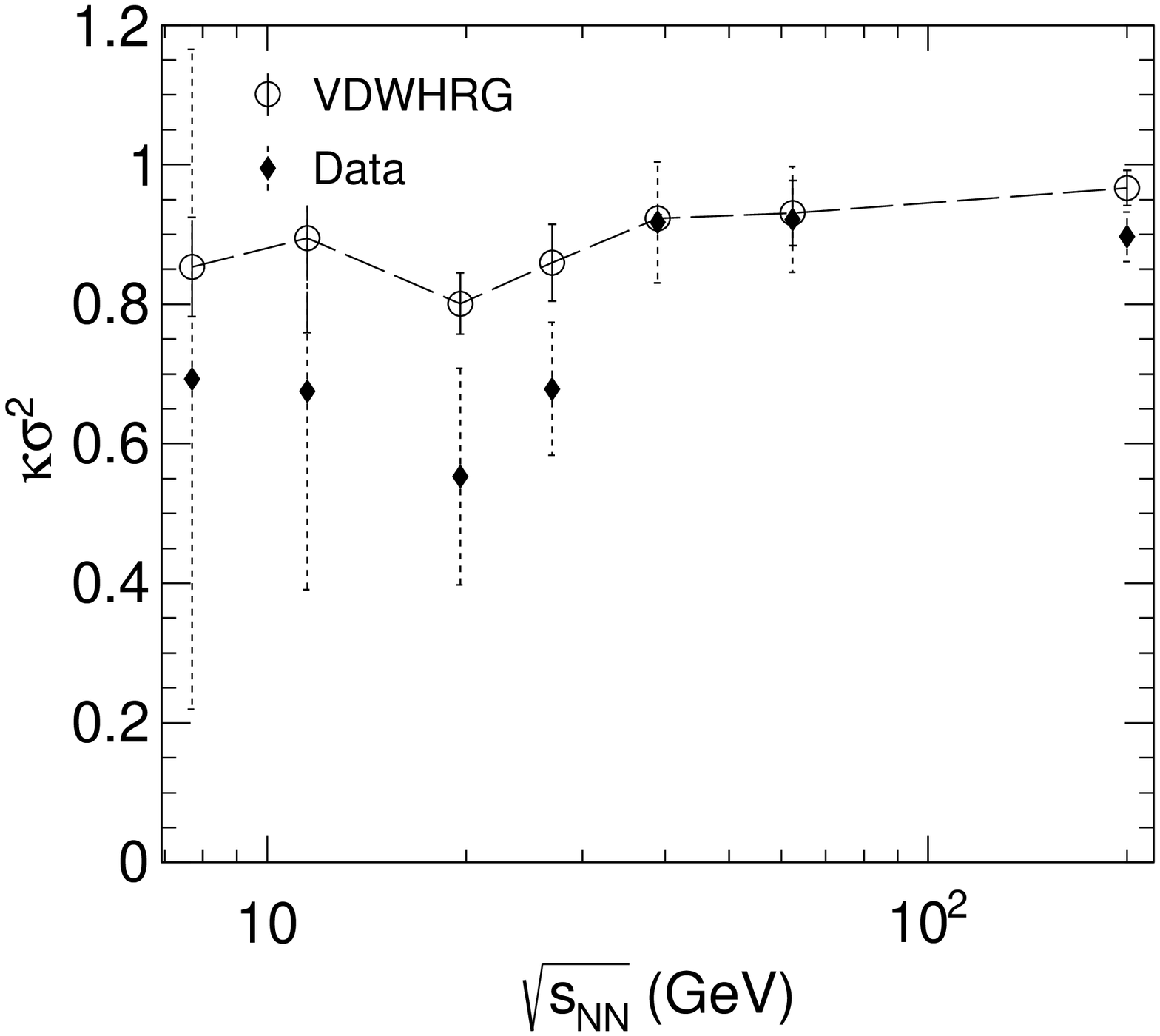}
 \caption{ $\sigma^2/M$, $S\sigma$ and $\kappa\sigma^2$ of net-proton
 in the VDWHRG model with the parameter Set-1. Results are compared
 with the experimental data of net-proton fluctuations for $0-5 \%$
 Au + Au collisions measured by the STAR
collaboration at RHIC~\cite{Adamczyk:2013dal}.  
Errors on data points are systematic and statistical errors added in quadrature and 
errors on model results are obtained by considering the values at $\chi^2_{min}+1$.}
 \label{fig:ratio_cum_all}
 \end{figure*}

Experimentally fluctuations of conserved charges are obtained by 
measuring the conserved number (net-charge or net-baryon or
net-strangeness) on the event-by-event basis within a certain
rapidity $y$ and transverse momentum $p_T$ acceptance. 
The net-number of the conserved quantity takes different values for
each event and hence gives a distribution when measured for a large
number of events. Mean of the distribution is the event average of the net-number
of the conserved charge. i.e.,
\begin{equation}
M_q=\left\langle N_q \right \rangle.
\end{equation}
The $\mathrm {n^{th}}$ order central moment is defined as
\begin{equation}
 \delta{N_q}^n = \left\langle \left( N_q - \left\langle N_q \right\rangle \right)^n \right\rangle.
\end{equation}
The mean ($M_q$), variance
($\sigma_q^2$), skewness ($S_q$) and kurtosis ($\kappa_q$) of distribution
of the conserved charge
are related to the central moments of the distribution and also to
different order of the corresponding susceptibilities by
the following relations:
\begin{equation}\label{eq:mean}
M_q = VT^3\chi_q^1,
\end{equation}
\begin{equation}\label{eq:variance}
 \sigma_q^2=\left\langle(\delta{ N_q})^2\right\rangle=VT^3\chi_q^2,
\end{equation}
\begin{equation}\label{eq:skewness}
 S_q=\frac{\left\langle(\delta{ N_q})^3\right\rangle}{\sigma_q^3}=\frac{VT^3\chi_q^3}{(VT^3\chi_q^2)^{3/2}},
\end{equation}
\begin{equation}\label{eq:kurtosis}
 \kappa_q=\frac{\left\langle(\delta{ N_q})^4\right\rangle}{\sigma_q^4}-3=\frac{VT^3\chi_q^4}{(VT^3\chi_q^2)^2}.
\end{equation}
The mean, variance, skewness are respectively estimations of the most
probable value, width, asymmetry and the peakedness of the
distribution, respectively.
The kurtosis indicates the sharpness of a distribution compared 
with the Gaussian distribution (for which all the moments higher than
second order are zero).
From Eqs. \ref{eq:mean} - \ref{eq:kurtosis}, volume independent ratios can be defined as:
\begin{subequations}
\label{allequations}
\begin{eqnarray}
&\sigma_q^2/M_q = \chi_q^2/\chi_q^1,\label{equationa}
\\
&S_q \sigma_q = \chi_q^3/\chi_q^2,\label{equationb}
\\
&\kappa_q \sigma_q^2 = \chi_q^4/\chi_q^2\label{equationc}.
\end{eqnarray}
\end{subequations}
The left-hand side quantities in Eqs. \ref{allequations} can be measured in the experiments while
the right hand-side quantities can be calculated in models like
VDWHRG. As already mentioned, non-monotonic variations of these quantities  with
beam energy ($\sqrt{\mathrm {s_{NN}}}$) are believed to be good signatures of a 
phase transition and CP.
The STAR collaboration has published all of the above-mentioned
observables for the net-proton, net-charge and net-kaon distributions at different energies 
ranging from $7.7$ GeV to $200$ GeV and at various centralities
\cite{Adamczyk:2013dal, Adamczyk:2014fia, Adamczyk:2017wsl}.
Similar observables for net-charge have been reported by
the PHENIX collaboration~\cite{Adare:2015aqk}. For the current study, 
we use only the published results for the net-proton number
distribution in central Au-Au collisions
from the STAR experiment and not the
subsequent preliminary unpublished results (as these could be subject
to changes). We have not considered the net-kaon and
net-charge results as they have large uncertainties and
the later has dominant contributions from resonance decay
effects~\cite{Garg:2013ata}. We have also not used the net-charge results from the 
PHENIX experiment as the acceptance for those
measurements ($|y| < 0.35$) are much smaller compared to STAR ($|y| < 0.5$). 
Studies suggest, to capture the relevant physics processes like CP it is ideally required
to have acceptance of at least 1 unit in rapidity~\cite{Jeon:2000wg}. Further, it
has been suggested that net-proton number is a sensitive observable
for CP physics~\cite{Hatta:2003wn}.

\section{Extraction of van der Waals parameters from experimental data}\label{Sc:Extraction}

\begin{figure*}
\centering
 \includegraphics[width=0.32\textwidth]{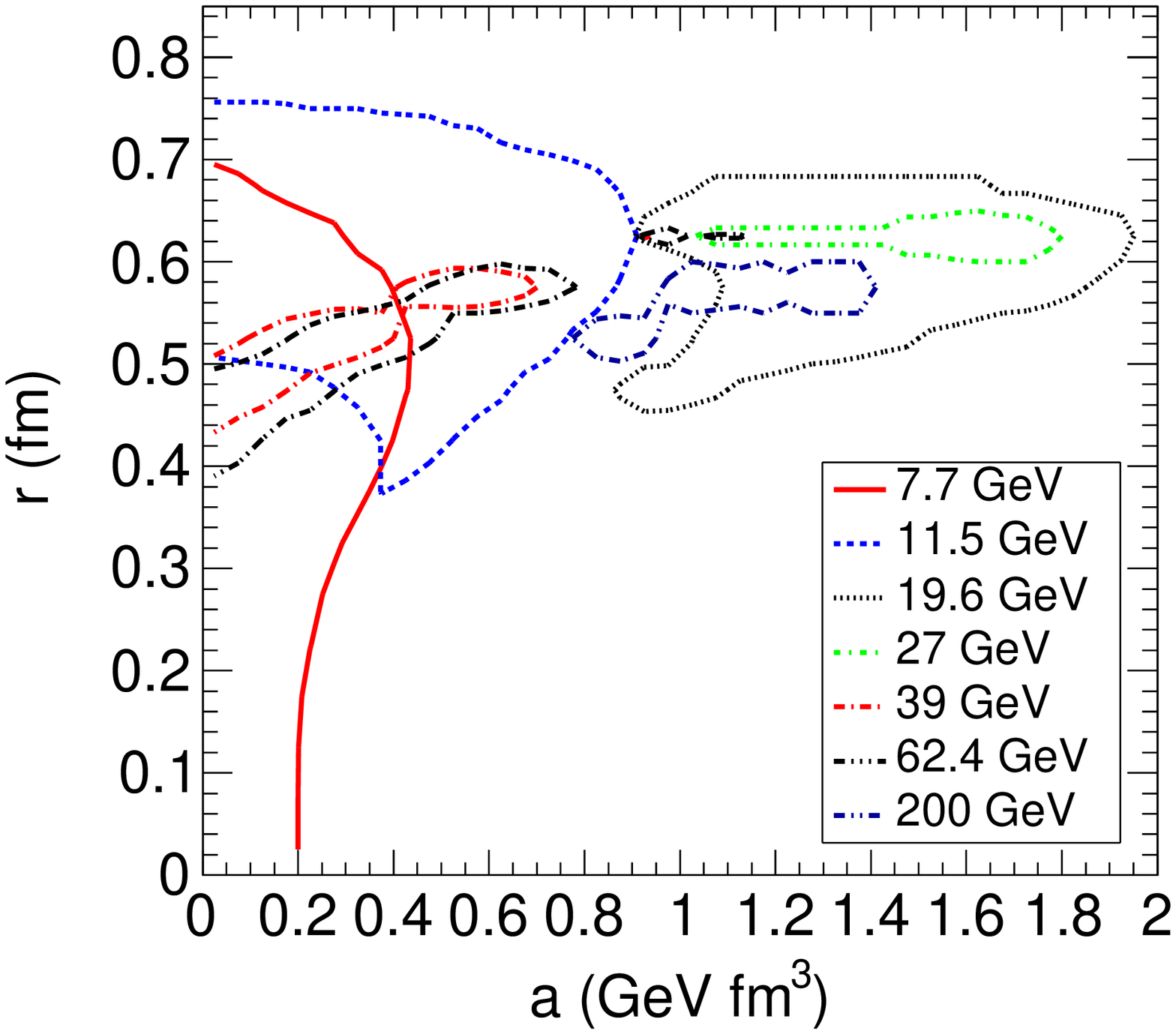}
 \includegraphics[width=0.32\textwidth]{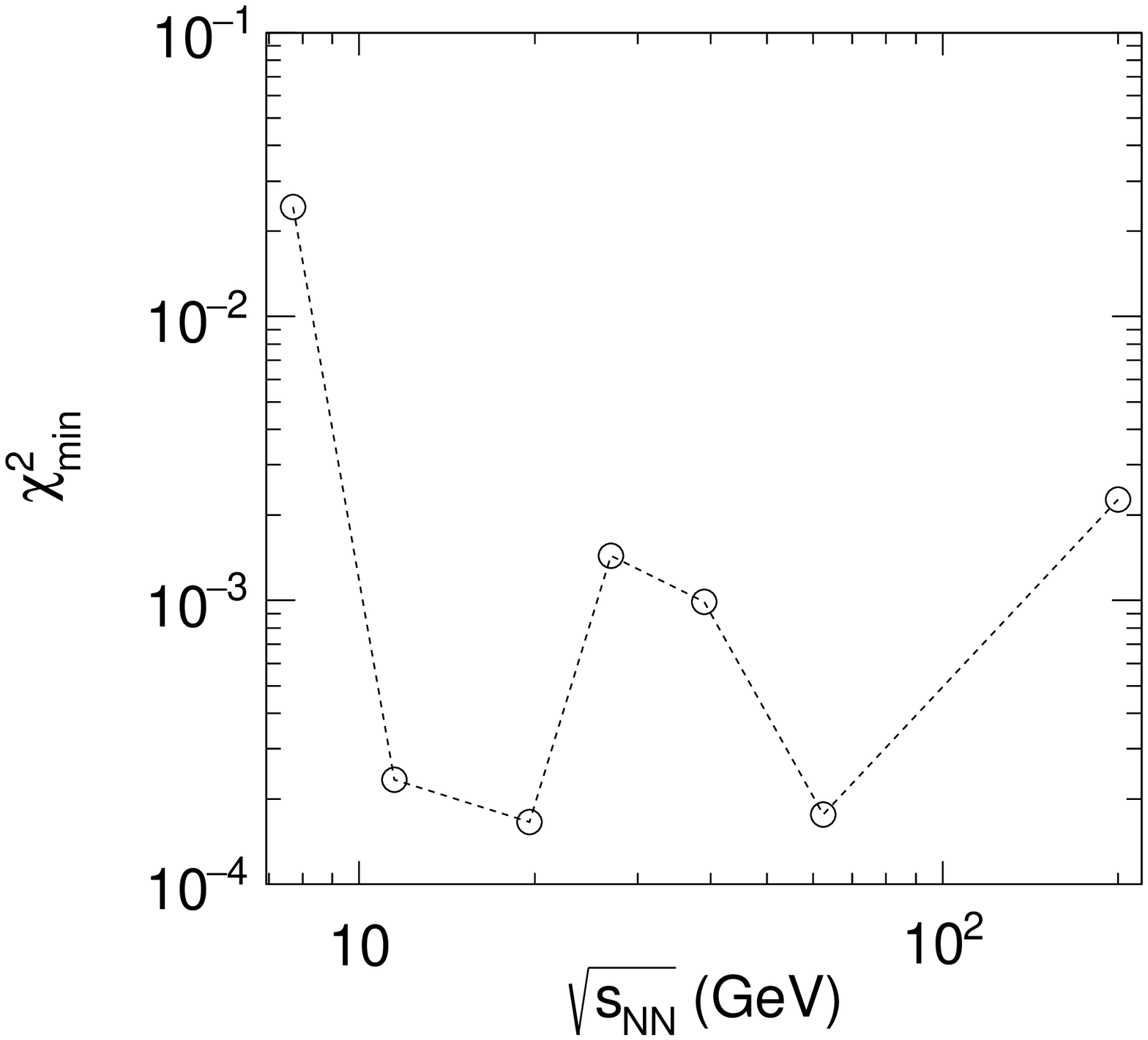}
 \caption{(Left) $1 \sigma$ contours of the van der Waals parameter 
 in $r,a$ plane at different $\sqrt{\mathrm {s_{NN}}}$ 
extracted from  $S\sigma$ and $\kappa\sigma^2$ of net-proton
(Set-2), (Right) minima of $\chi^2$ at different $\sqrt{\mathrm {s_{NN}}}$.}
\label{fig:a_r_3242}
 \end{figure*}

\begin{figure*}
\centering
 \includegraphics[width=0.32\textwidth]{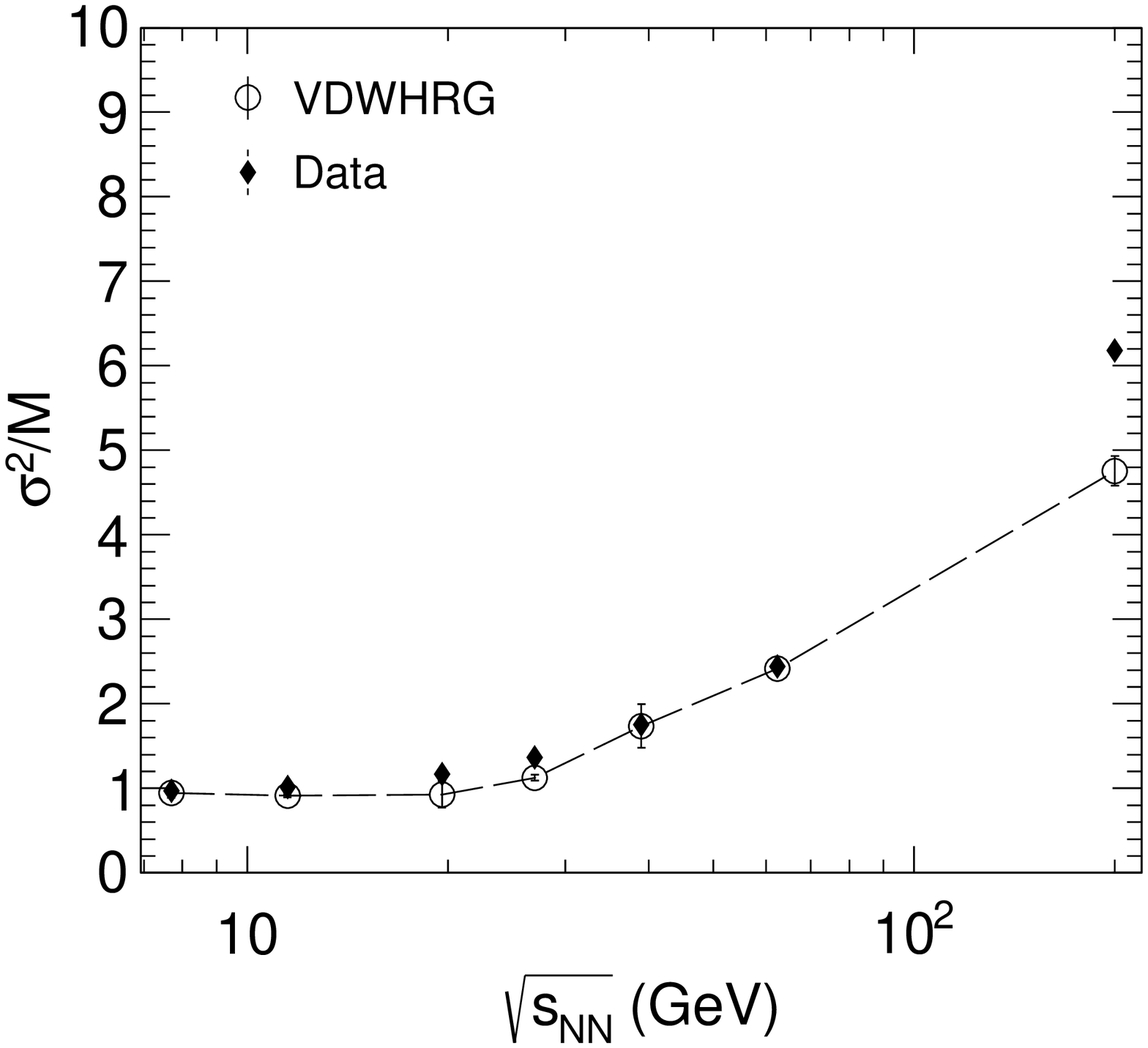}
 \includegraphics[width=0.32\textwidth]{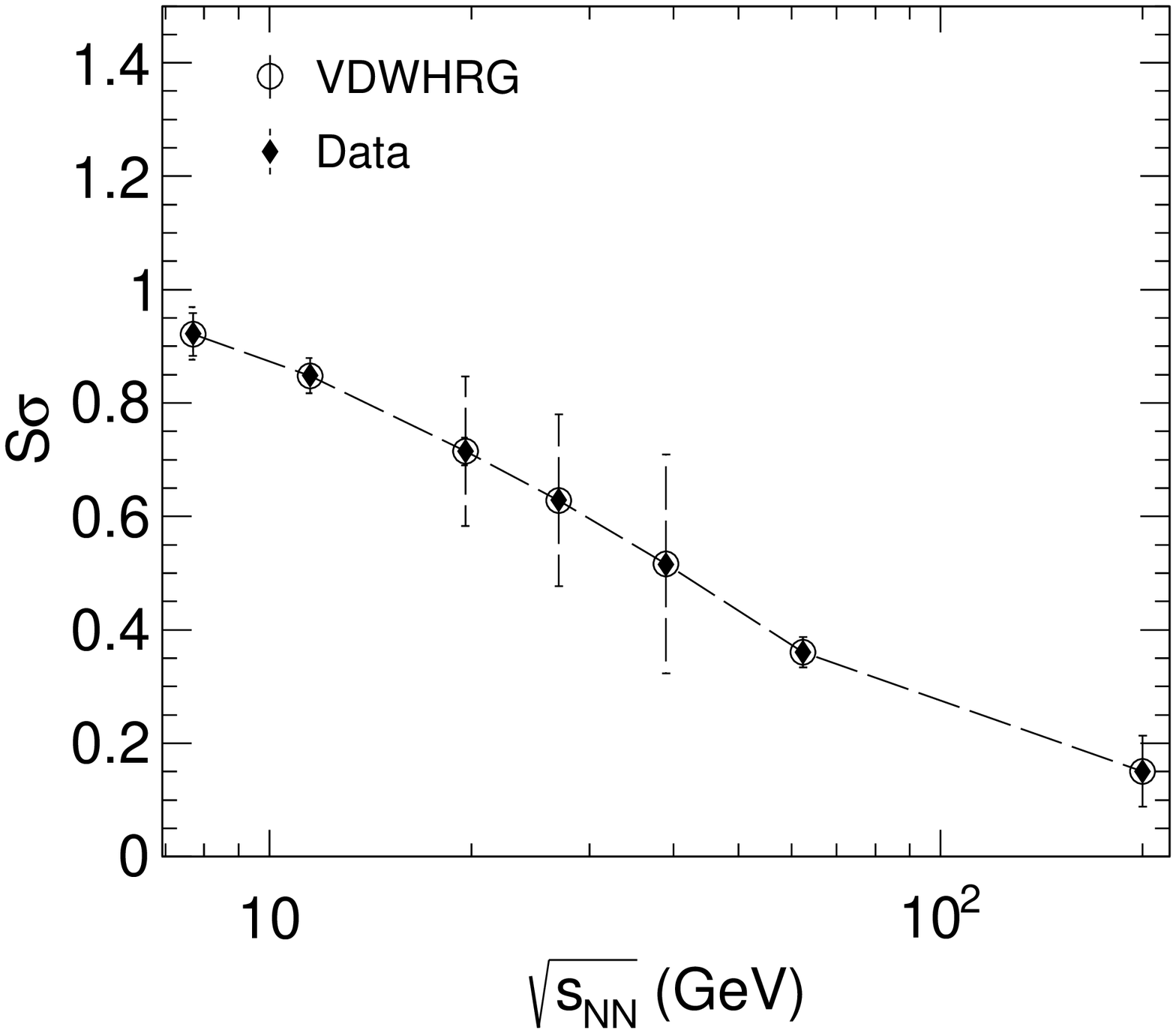}
 \includegraphics[width=0.32\textwidth]{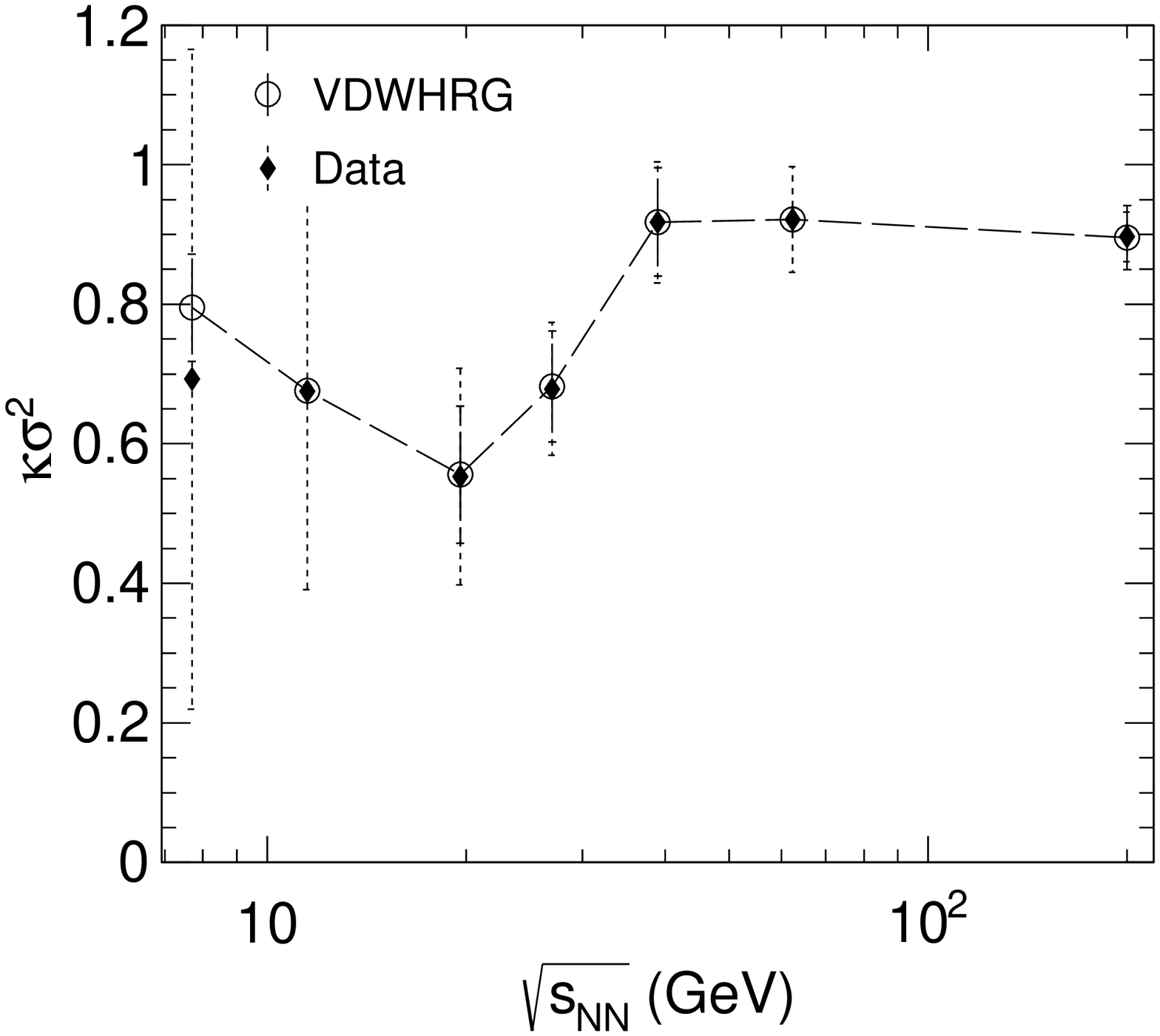}
 \caption{ Same as Fig.~\ref{fig:ratio_cum_all} but for the 
 parameter Set-2.}
 \label{fig:ratio_cum_3242}
 \end{figure*}

In this work, we have extracted the attractive and repulsive 
parameters, $a$ and $r$, of the 
VDWHRG model from the fluctuation observables of net protons measured
by the STAR collaboration~\cite{Adamczyk:2013dal}.
The $T$ and the $\mu_B$ at the chemical
freeze-out can be parametrized by the following functions~\cite{Cleymans:2005xv}:
\begin{equation}
 T (\mu_B) = a - b \mu_B^2 -c \mu_B^4,
\end{equation}
where $a = (0.166 \pm 0.002)$ GeV, $b = (0.139 \pm 0.016)~\text{GeV}^{-1}$,
$c = (0.053 \pm 0.021)~\text{GeV}^{-3}$ and
\begin{equation}
 \mu_B(\sqrt{\mathrm {s_{NN}}}) = \frac{d}{1 + e \sqrt{\mathrm {s_{NN}}}},
\end{equation}
where $d = 1.308 \pm 0.028~\text{GeV}$ and $e = 0.273 \pm 0.008 ~\text{GeV}^{-1}$.
These chemical freeze-out parameters provide
a good quantitative description of the hadronic yields over a wide
range of $\sqrt{\mathrm {s_{NN}}}$. At different $\sqrt{\mathrm {s_{NN}}}$ we have used
the chemical freeze-out $T$ and $\mu_B$ from the above mentioned parametrized equations.
To extract the van der Waals parameters $a$ and $r$
at a particular $\sqrt{\mathrm {s_{NN}}}$ we have used a $\chi^2$ minimization
technique where $\chi^2$ is defined as
\begin{equation}
\chi^2 = \frac{1}{N} \sum_i^N \frac{\left( O_i^{expt}-O_i^{model} \right)}{\left(\sigma_i^{expt}\right)^2}
\end{equation}
where $O_i^{model}$ is the $\mathrm {i^{th}}$ observable calculated in model
whereas $O_i^{expt}$ and $\sigma_i^{expt}$ are its 
experimental value and uncertainty in the measurement respectively. 
For the experimental uncertainties, we
have quadratically added the statistical and the systematic errors for the observables.
$N$ in the above equation is the number of observables
used to calculate $\chi^2$. The $1 \sigma$ error of the extracted parameters
correspond to $\chi^2_{min} + 1$. 

We have to extract values of two parameters and we have three experimental
observables $\sigma^2/M, S \sigma$ and $\kappa \sigma^2$ of net-proton
distribution for 7 different $\sqrt{\mathrm {s_{NN}}}$. Two sets of van
der Waals parameters ($a$ and $r$) have been extracted by fitting the fluctuation observables
in our present analysis which are listed in the Table.~\ref{tableLabel1}.
For the first set  (Set-1), we use $\sigma^2/M, S \sigma$ and 
$\kappa \sigma^2$ of net-proton distributions at 0-5 \% centrality in Au + Au collisions measured by the STAR
Collaboration at RHIC~\cite{Adamczyk:2013dal}. For the second set
(Set-2) we have 
used the $S \sigma$ and $\kappa \sigma^2$ data. The $S \sigma$ and $\kappa \sigma^2$
are common for both the sets. It may be noted that these two
observables are expected to be more sensitive to the CP physics~\cite{Stephanov:2008qz}. 
Net-proton fluctuations were experimentally measured at
mid-rapidity ($|y| < 0.5$) and within the transverse momentum
range $0.4 < p_T < 0.8$ GeV. To incorporate acceptance
cuts in our model, we have written the
$d^3p$ and the single particle energy $E$ as
$d^3p = 2\pi~p_T~m_T~\cosh~y~dy~dp_T$,
and $E = m_T~\cosh y$ where $m_T = \sqrt{p_T^2 +m^2}$.
Then the interaction ranges in $y$ and $p_T$ are chosen 
as $-0.5$ to 0.5 and 0.4 to 0.8 GeV respectively to make
the model results comparable with the experimental data.

The left plot of Fig.~\ref{fig:a_r_all} shows the $1 \sigma$ contours of the 
van der Waals attractive and repulsive parameters, $a, r$, for Set-1 at 
different $\sqrt{\mathrm {s_{NN}}}$ from 7.7 GeV up to 200 GeV.
We find that both $a$ and $r$ shows a possibility of reaching maxima at
$\sqrt{\mathrm {s_{NN}}} = 19.6$ GeV. It may be noted that the experimental data of $\kappa
\sigma^2$ shows a minimum around this energy. We also find that the repulsive
parameter strength increases as we go down from $\sqrt{\mathrm {s_{NN}}} = 200$ GeV
to 19.6 GeV. The attractive parameter strength ranges are of similar
order for these energies. However for $\sqrt{\mathrm {s_{NN}}}$ = 7.7 GeV to 11.5
GeV both the strengths of $a$ and $r$ shows a decrease relative to
19.6 GeV. 
Right plot of Fig.~\ref{fig:a_r_all} shows the minima of the 
$\chi^2$ at different $\sqrt{\mathrm {s_{NN}}}$.

In Fig.~\ref{fig:ratio_cum_all} we show the variations of
$\sigma^2/M, S\sigma$ and $\kappa \sigma^2$ from the VDWHRG model using the parameter obtained by
fitting the net-proton experimental data Set-1. We also compare the model results with the experimental
data. We observe that with this parameter set, VDWHRG model can
describe $\sigma^2/M$ and $S\sigma$ at all the energies within the
uncertainties.
In addition, the measured $\kappa \sigma^2$ of net-proton distribution can also be described 
qualitatively. At $\sqrt{\mathrm {s_{NN}}}= $ 19.6 and 27 GeV, VDWHRG model with
parameters obtained using the data in Set-1 slightly overestimate the
experimental values of $\kappa \sigma^2$ and at all other energies
model and data agree within the uncertainties.

$1 \sigma$ contours of the VDW parameters using the data in Set-2 for different $\sqrt{\mathrm {s_{NN}}}$ are
shown in left plot of Fig.~\ref{fig:a_r_3242}.  The data in Set-2 are more
sensitive to CP physics. Here we observe that the attractive VDW
parameter strength becomes large for $\sqrt{\mathrm {s_{NN}}}$ = 19.6 and 27
GeV relative to the values obtained at  $\sqrt{\mathrm {s_{NN}}}$ = 7.7, 11.5,
39, 62.4 GeV. 
The values of the attractive VDW parameter in Set-2 are relatively larger compared to Set-1.
Particularly, at $\sqrt{\mathrm {s_{NN}}} = 19.6$ GeV, value of $a$ in Set-2 is
double compared to that for Set-1. The larger values of $a$ for the Set-2
indicates that, the higher order fluctuation observables, $S\sigma$ and $\kappa \sigma^2$ are more
sensitive to the attractive interaction than $\sigma^2/M$.
Further, large increase of the attractive interaction near $19.6$ GeV
might be due to the existence of a physics processes where attractive
interactions are dominant such as a CP.
The repulsive VDW parameter varies approximately within 0.4 to 0.75 fm at all the $\sqrt{\mathrm {s_{NN}}}$
except $7.7$ GeV where the value $r$  is not well constrained due to
large uncertainties in the measurement. 
The maxima of $r$ observed at $\sqrt{\mathrm {s_{NN}}} = 19.6$ GeV in Set-2 is
comparable to that from the Set-1.
This indicates that sensitive to the  repulsive interaction is of the same order for 
the various choice of observables. 
Right plot of Fig.~\ref{fig:a_r_3242} shows the minima of the 
$\chi^2$ at different $\sqrt{\mathrm {s_{NN}}}$.

Figure~\ref{fig:ratio_cum_3242} shows the variations of
$\sigma^2/M, S\sigma$ and $\kappa \sigma^2$ of net-proton in VDWHRG
model using the parameters obtained from the fit to data given in Set-2.
We compare the model results with the experimental data. 
It can be seen that, with this parameter set, $\sigma^2/M$ of 
net-proton distribution can be described in all the energies except for 200 GeV
where model underestimates the data. 
Further, this parameter set can describe $S\sigma$ and $\kappa \sigma^2$
of net-proton distribution measured in all the energies from 7.7 to 200 GeV.

\section{Summary}\label{Sc:Summary}
We have extracted the van der Waals attractive and repulsive
parameters, $a$ and $r$, by fitting the experimental data of fluctuation
observables of net protons measured by the STAR Collaboration using VDWHRG 
model. The variation of VDW parameters reflective of the attractive and
repulsive nature of the interactions for the QCD matter produced in
heavy-ion collisions with collision energy has been discussed. 
Two sets of ($a, r$) parameters have been extracted. In the
first set (Set-1), we have used $\sigma^2/M, S\sigma$ and $\kappa \sigma^2$
of net-proton. In another set (Set-2) we have used only  $S\sigma$ and $\kappa \sigma^2$
of net-proton. We have incorporated the proper experimental acceptances in our 
calculation.  We have observed that the higher order fluctuation observables $S\sigma$
and $\kappa \sigma^2$ are more sensitive to the attractive interaction.
Large increases in the attractive interaction (i.e., $a$) 
is observed near $\sqrt{\mathrm {s_{NN}}} = 19.6$ GeV, indicates change in
equation-of-state near this energy. Further, we observe that with the
parameter Set-1, VDWHRG model can describe 
$\sigma^2/M$ and $S\sigma$ of net-proton in all the energies studied
in this work. The $\kappa \sigma^2$ of net-proton can also
be described qualitatively. On the other hand, with the parameter Set-2,
VDWHRG model can describe $\sigma^2/M$ of net-proton from 
$\sqrt{\mathrm {s_{NN}}} = 7.7$ to 62.4 GeV. Not only that,
the $S\sigma$ and $\kappa \sigma^2$ of
net-proton can be described within the error bars
by VDWHRG model in all the energies.

\section*{Acknowledgement}
SS acknowledges financial support from DAE, Government of India. BM
acknowledges support from DST through J. C. Bose Fellowship for this work.

\bibliography{RefFile}

%

\end{document}